# On "P=NP Linear programming formulation of the Traveling Salesman Problem": A Reply to Hofman's Claim of "Counter-Example"

Moustapha Diaby, 2006

## 1. Overview of Hofman's "Constructs":

### 1.1. Hofman's Concept of "Valleys"

(Paraphrasing Hofman)

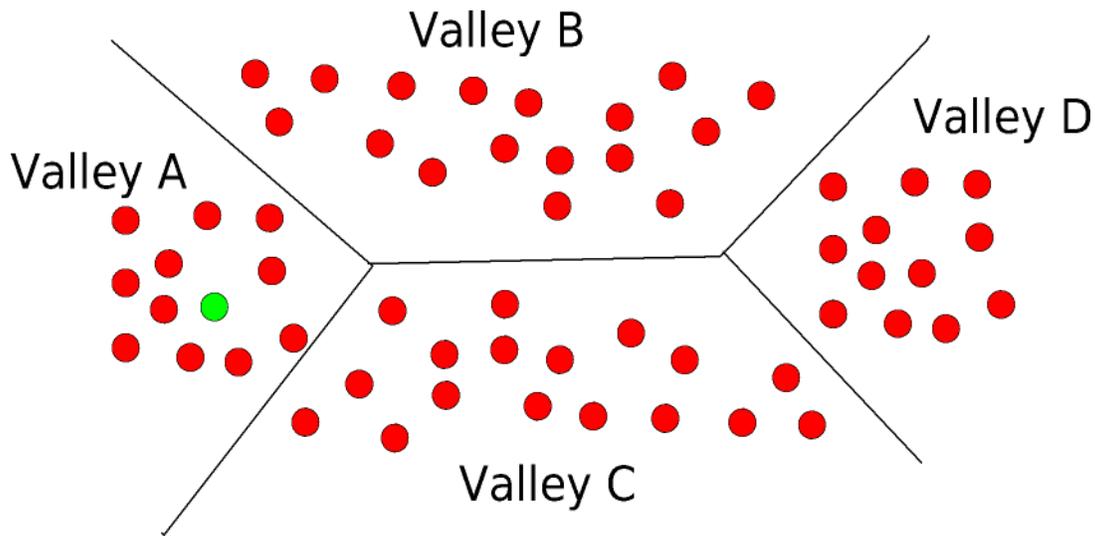

*Figure 3 Four valleys*

Imagine *v* valleys and all cities placed in this valleys. Cost of traveling from town to town in valley is very small (negligible), let us say equal to 1. When salesmen wants to go to other valley he has to pass mountain chain – this is of a huge cost, let us say equal to 1000. Salesman lives in one of towns in valley A. Optimal solution must then contain 4 mountain chains crossing (from A to B, B to C, C to D and D to A), its overall cost is then 4*1000+*X* (*X* is cost of in valley traveling and is very small).

### 1.2. Hofman's Principle for Constructing "Solutions"

1. visit once all cities in a "valley"
2. fractions of salesman "travel" to other "valleys"
3. visit **multiple times** every city in valleys in which fractions of salesman is "sent"
4. Link-up the "fractional salesmen" in yet another "valley"
5. visit all towns in that valley
6. return to starting town
7. Super-impose different set of "solutions" obtained using 1) – 6) above to create an overall solution that is feasible to the model in Diaby

For example, according to Hofman, super-imposition of "solutions" such as illustrated in Figures 4, 5, 6 below, leads to the full "solution" shown Figure 13 that is purported to be feasible to the model in Diaby.

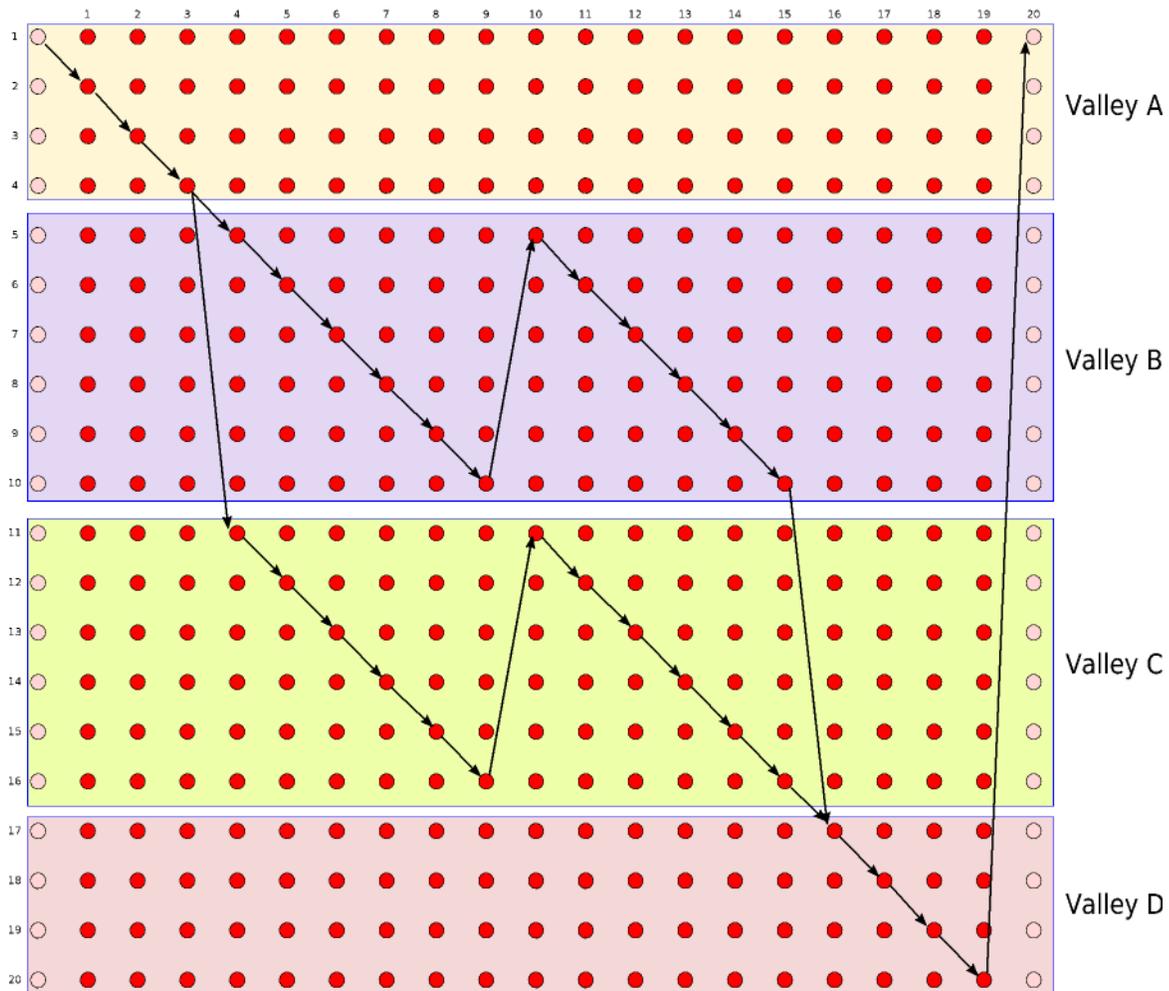

*Figure 4 Four valleys path on diagram*

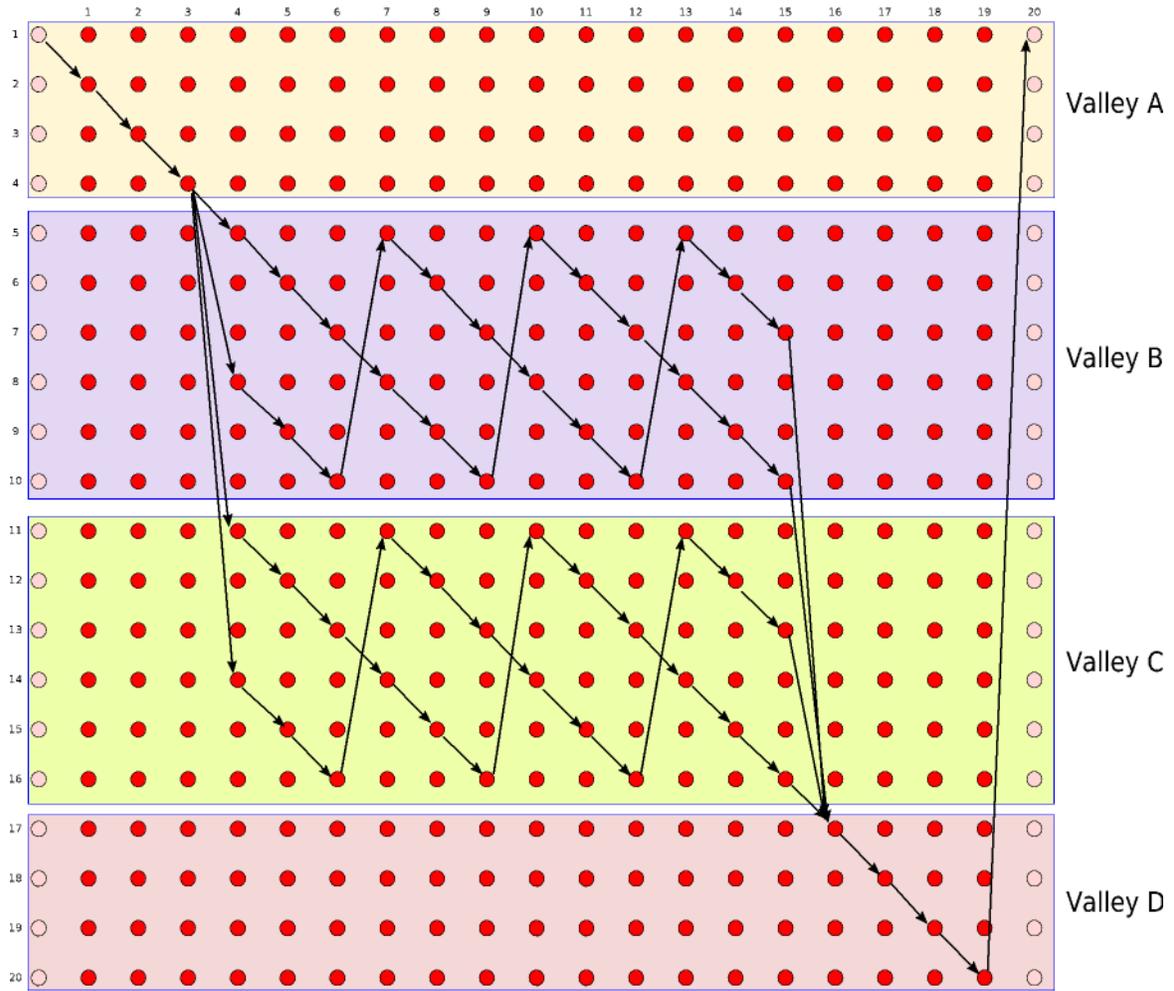

*Figure 5 Four valleys path on diagram – first*

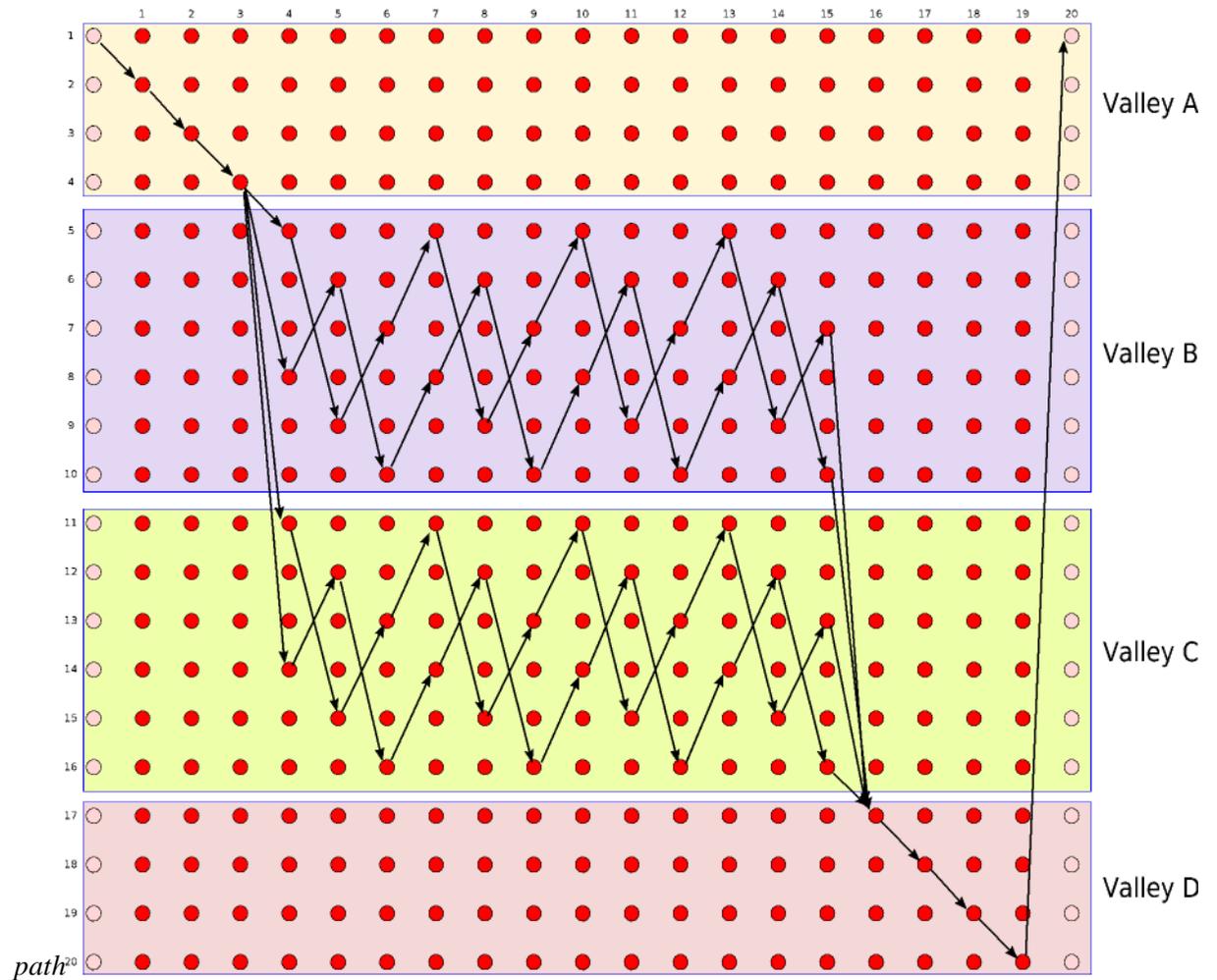

*path*

*Figure 6 Four valleys path on diagram – additional paths*

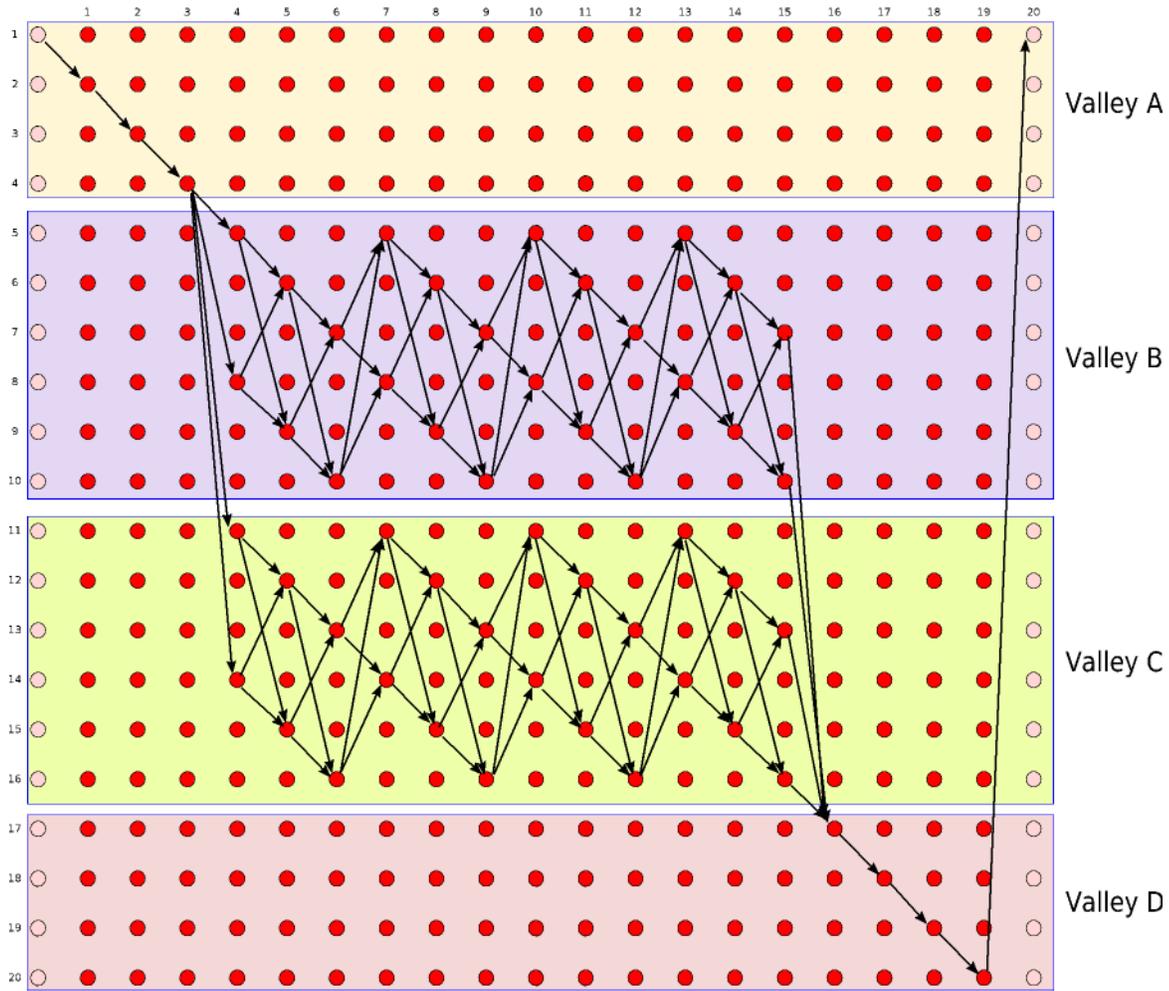

*Figure 7 All necessary paths for four valleys*

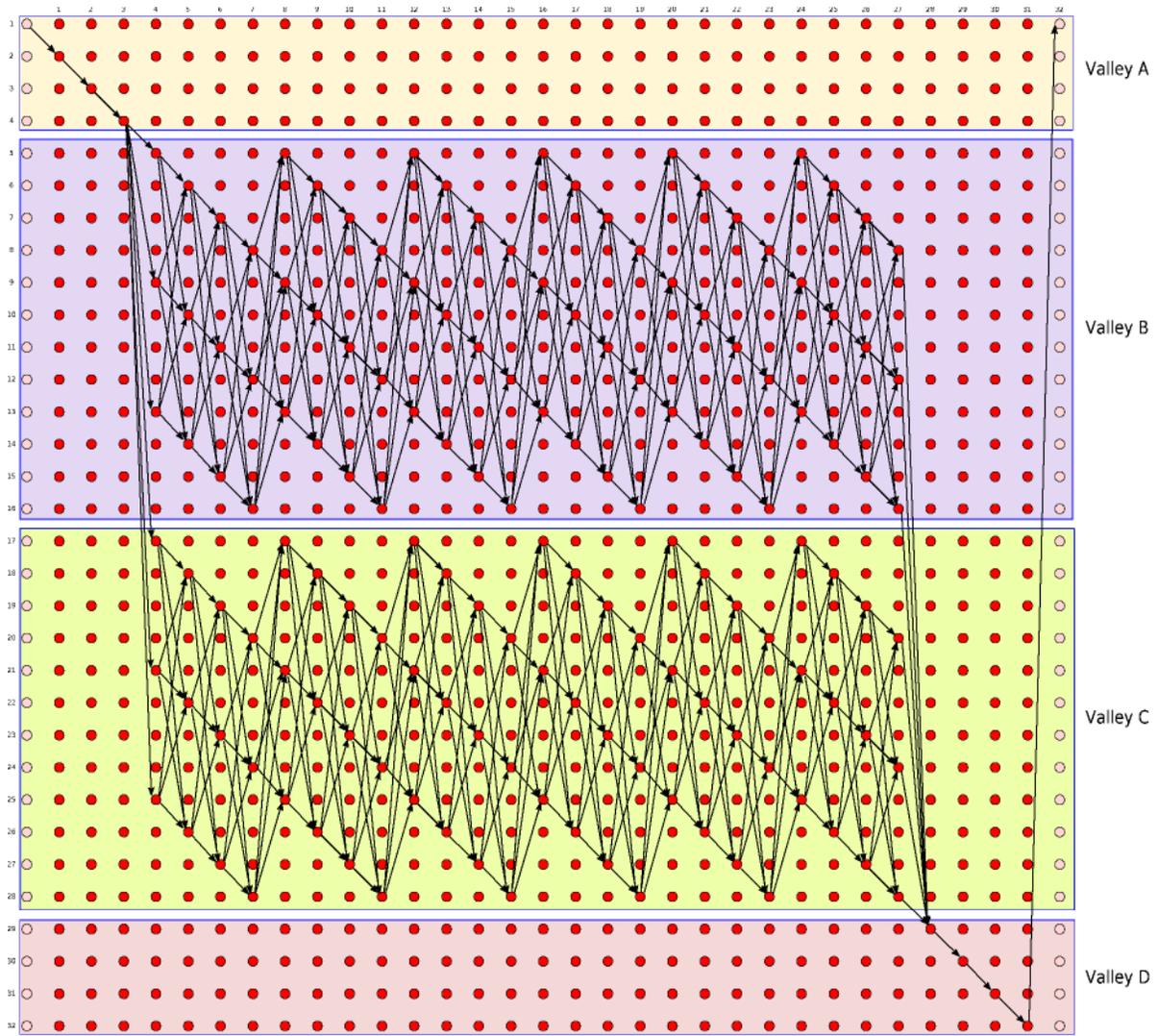

*Figure 13 All necessary paths for four valleys*

**2. Flaws in Hofman's "Constructs"**

A major flaw in Hofman's "construction" scheme is that it overlooks the **impossibility** of having the "salesman" (be it/he/she "full" or fractional) visit n cities in more than n "travels" without visiting any city more than once. This impossibility holds true in my model because of Propositions 2 and 3 in the paper. Furthermore, because of this impossibility, the flows within each of the "valleys" where Hofman sends "fractional-salesmen" cannot be feasible for my model.

For example, consider arc (5,4,6) in Hofman's "valley" B in Figure 13. It is impossible to have a set of positive $y_{5,4,6,i_s,s,j_s}$ for s = 4, 5, …, 23 along with positive $z_{5,4,6,i_s,s,j_s,i_t,t,j_t}$ for s = 5, …, 22, and t = s+1, …, 23 such that all the constraints of my model (or Proposition 2) are satisfied.  In other words, it is impossible for the flow from arc (5,4,6) to propagate onto arcs onto arcs at stage 23 without violations of constraints 2.14 of my model. The required "travel" involves 19 stages. Since there are only 6 cities included in Hofman's "valley" B, some of those cities must be "visited" more than once.